\documentclass[a4paper,12pt]{article}%
\usepackage{amsmath}
\usepackage{amsfonts}
\usepackage{amssymb}
\begin{document}

\title{\textsf{New BPS Solitons in 2+1 Dimensional Noncommutative } $CP^{1}$
\textsf{Model}}
\author{\textsf{Hideharu Otsu } \thanks{otsu@vega.aichi-u.ac.jp}\\Faculty of Economics, \\Aichi University, Toyohashi, Aichi 441-8522, Japan
\and \textsf{Toshiro Sato} \thanks{tsato@matsusaka-u.ac.jp}\\Faculty of Policy Science, Matsusaka University, \\Matsusaka, Mie 515-8511, Japan
\and \textsf{Hitoshi Ikemori } \thanks{ikemori@biwako.shiga-u.ac.jp}\\Faculty of Economics, Shiga University, \\Hikone, Shiga 522-8522, Japan
\and \textsf{Shinsaku Kitakado }\thanks{kitakado@ccmfs.meijo-u.ac.jp}\\Department of Information Sciences, Meijo University, \\Tempaku, Nagoya 486-8502, Japan }
\date{}
\maketitle

\begin{abstract}
Investigating the solitons in the non-commutative $CP^{1}$ model, we have
found a new set of BPS solitons which does not have counterparts in the
commutative model.

\end{abstract}

\newpage

\section[ Introduction]{Introduction}

\smallskip Field theories on the non-commutative space have been extensively
studied in the last few years. Particularly, BPS solitons are interesting,
because they might not share the common features with those on the commutative
space\cite{Harvey:2001yn}\cite{Nekrasov:1998ss}\cite{Gopakumar:2000zd}.

Solitons in the non-commutative $CP^{1}$ model have been studied in
\cite{Lee:2000ey} and further developed in \cite{Furuta:2002ty} in connection
with the dynamical aspects of the theory. The non-BPS solitons, that do not
exist in the commutative case, have been studied in \cite{Furuta:2002nv}.
These investigations were reviewed in \cite{Murugan:2002rz}.

In this paper, we report on a set of new BPS solitons in the non-commutative
$CP^{1}$ model, that does not exist in the commutative limit.

We consider the $CP^{1\text{ }}$model on 2+1 dimensional non-commutative
spacetime. The space coordinates obey the commutation relation%
\begin{equation}
\left[  x,y\right]  =i\theta,
\end{equation}
or%

\begin{equation}
\left[  z,\bar{z}\right]  =\theta>0,
\end{equation}
in terms of the complex variables, $z=\frac{1}{\sqrt{2}}(x+iy)$ and $\bar
{z}=\frac{1}{\sqrt{2}}(x-iy)$. The Hilbert space can be described in terms of
the energy eigenstates $\left\vert n\right\rangle $ of the harmonic oscillator
whose creation and annihilation operators are $\bar{z}$ and $z$ respectively,%
\begin{align}
z\left\vert n\right\rangle  &  =\sqrt{\theta n}\left\vert n-1\right\rangle
,\ \ \\
\bar{z}\left\vert n\right\rangle  &  =\sqrt{\theta(n+1)}\left\vert
n+1\right\rangle ,\nonumber
\end{align}
The $CP^{1}$ lagrangian is%
\begin{equation}
L=\text{$\mathrm{Tr}$}(\left\vert D_{t}\Phi\right\vert ^{2}-\left\vert
D_{z}\Phi\right\vert ^{2}-\left\vert D_{\bar{z}}\Phi\right\vert ^{2}),
\end{equation}
where $\Phi$ is a 2-component complex vector with the constraint
$\Phi^{\dagger}\Phi=1$. We consider $\Phi$ to be the fundamental field and
thus to be non-singular. Tr denotes the trace over the Hilbert space
as\smallskip%
\begin{equation}
\text{$\mathrm{Tr}$}\mathcal{O}=2\pi\theta\sum_{n=0}^{\infty}\left\langle
n\right\vert \mathcal{O}\left\vert n\right\rangle .
\end{equation}
The covariant derivative is defined by
\begin{equation}
D_{a}\Phi=\partial_{a}\Phi-i\Phi A_{a},\quad A_{a}=-i\Phi^{\dagger}%
\partial_{a}\Phi,\quad\left(  a=t,z,\bar{z}\right)  ,
\end{equation}
where $\partial_{z}=-\theta^{-1}\left[  \bar{z},\ \right]  $ and
$\partial_{\bar{z}}=\theta^{-1}\left[  z,\ \right]  $.

For the static configurations, topological charge and static energy are given
by
\begin{equation}
Q=\frac{1}{2\pi}\mathrm{Tr}\left(  \left\vert D_{z}\Phi\right\vert
^{2}-\left\vert D_{\bar{z}}\Phi\right\vert ^{2}\right)  , \label{Q1}%
\end{equation}
and%
\begin{equation}
E=\mathrm{Tr}\left(  \left\vert D_{z}\Phi\right\vert ^{2}+\left\vert
D_{\bar{z}}\Phi\right\vert ^{2}\right)  \geq2\pi\left\vert Q\right\vert .
\label{E1}%
\end{equation}
The configuration which saturates the energy bound satisfies the BPS soliton
equation
\begin{equation}
D_{\bar{z}}\Phi=0, \label{bps1-1}%
\end{equation}
or BPS anti-soliton equation%
\begin{equation}
D_{z}\Phi=0. \label{bps2-1}%
\end{equation}

\smallskip It is convenient \cite{Lee:2000ey} to introduce the 2-component
complex vector $W$ and the projection operator $P$ as
\begin{equation}
\Phi=W\frac{1}{\sqrt{W^{\dagger}W}},\ \ P=\Phi\Phi^{\dagger}.
\label{Projection}%
\end{equation}
In terms of the projection operator, BPS soliton equations (\ref{bps1-1}) and
(\ref{bps2-1}) are \cite{Hadasz:2001cn}\cite{Lechtenfeld:2001aw}%
\cite{Gopakumar:2001yw}
\begin{equation}
\left(  1-P\right)  zP=0, \label{bps1-2}%
\end{equation}
and%
\begin{equation}
\left(  1-P\right)  \bar{z}P=0, \label{bps2-2}%
\end{equation}
which indicate respectively%
\begin{equation}
zW=WV, \label{bps1-3}%
\end{equation}
and%
\begin{equation}
\bar{z}W=WV, \label{bps2-3}%
\end{equation}
where $V$ is a scalar function \cite{Lee:2000ey}. Topological charge
(\ref{Q1}) and static energy (\ref{E1}) can be expressed as
\begin{equation}
Q=\frac{1}{2\pi}\text{Tr}\left\{  \partial_{\bar{z}}\Phi^{\dagger}\left(
1-P\right)  \partial_{z}\Phi-\partial_{z}\Phi^{\dagger}\left(  1-P\right)
\partial_{\bar{z}}\Phi\right\}  , \label{Q2}%
\end{equation}
and%
\begin{equation}
E=\text{Tr}\left\{  \partial_{\bar{z}}\Phi^{\dagger}\left(  1-P\right)
\partial_{z}\Phi+\partial_{z}\Phi^{\dagger}\left(  1-P\right)  \partial
_{\bar{z}}\Phi\right\}  . \label{E2}%
\end{equation}

The examples of BPS soliton are $W=(z^{n},1)^{t}$ , with topological charge
$Q=n$ and energy $E=2\pi n$, and those of BPS anti-soliton are $W=(\bar{z}%
^{n},1)^{t}$ with topological charge $Q=-n$ and energy $E=2\pi n$
\cite{Lee:2000ey}. These configurations are solitons also in the commutative theory.

\section[ New Solitons]{New Solitons}

\smallskip We have found that the following is the BPS soliton solution of the
non-commutative $CP^{1}$ model. The configuration of the soliton with the
topological charge $Q=-n$ is%
\begin{equation}
\Phi=\left(
\begin{array}
[c]{c}%
\bar{z}^{n}\dfrac{1}{\sqrt{\prod_{l=1}^{n}(\bar{z}z+l\theta)}}\\
0
\end{array}
\right)  , \label{phi1}%
\end{equation}
which can also be expressed in terms of projection operator $P$ as%
\begin{equation}
P=\left(
\begin{array}
[c]{cc}%
1-\sum_{m=0}^{n-1}\left\vert m\right\rangle \left\langle m\right\vert  & 0\\
0 & 0
\end{array}
\right)  . \label{P1}%
\end{equation}
For $Q=n,$ on the other hand, the soliton can be written as%
\begin{equation}
\Phi=\left(
\begin{array}
[c]{c}%
\dfrac{1}{\sqrt{\prod_{l=1}^{n}(\bar{z}z+l\theta)}}z^{n}\\
\sum_{m=0}^{n-1}\left\vert m\right\rangle \left\langle m\right\vert
\end{array}
\right)  , \label{phi2}%
\end{equation}
and the corresponding projection operator expression is%
\begin{equation}
P=\left(
\begin{array}
[c]{cc}%
1 & 0\\
0 & \sum_{m=0}^{n-1}\left\vert m\right\rangle \left\langle m\right\vert
\end{array}
\right)  . \label{P2}%
\end{equation}
We can straightforwardly confirm that (\ref{P1}) and (\ref{P2}) satisfy the
BPS equations (\ref{bps2-2}) and (\ref{bps1-2}) respectively. The energy of
these solitons are of course $E=2\pi n$ due to the BPS property.

We calculate the topological charge of (\ref{phi1}) and (\ref{phi2}). For BPS
anti-solitons (\ref{phi1}), we use%
\begin{equation}
\partial_{\bar{z}}\Phi^{\dagger}\left(  1-P\right)  \partial_{z}\Phi=0
\label{as1}%
\end{equation}
which follows from (\ref{bps2-1}), and
\begin{align}
&  \partial_{z}\Phi^{\dagger}\left(  1-P\right)  \partial_{\bar{z}}%
\Phi\label{as2}\\
&  =\theta^{-2}\left[  \frac{1}{\sqrt{\prod_{l=1}^{n}(\bar{z}z+l\theta)}}%
z^{n}\ ,\ \bar{z}\right]  \left(  \sum_{m=0}^{n-1}\left\vert m\right\rangle
\left\langle m\right\vert \right)  \left[  z\ ,\ \bar{z}^{n}\frac{1}%
{\sqrt{\prod_{l=1}^{n}(\bar{z}z+l\theta)}}\right] \nonumber\\
&  =\theta^{-1}n\left\vert 0\right\rangle \left\langle 0\right\vert .\nonumber
\end{align}
Substituting (\ref{as1}) and (\ref{as2}) into (\ref{Q2}), the topological
charge is%
\begin{equation}
Q=\frac{1}{2\pi}2\pi\theta\sum_{k=0}^{\infty}\left\langle k\right\vert \left(
-\theta^{-1}n\left\vert 0\right\rangle \left\langle 0\right\vert \right)
\left\vert k\right\rangle =-n. \label{Q3}%
\end{equation}
Similarly, for BPS solitons (\ref{phi2}), we use%
\begin{equation}
\partial_{z}\Phi^{\dagger}\left(  1-P\right)  \partial_{\bar{z}}\Phi=0
\label{S2}%
\end{equation}
which follows from (\ref{bps1-1}), and%
\begin{align}
\partial_{\bar{z}}\Phi^{\dagger}\left(  1-P\right)  \partial_{z}\Phi &
=\theta^{-2}\left[  \sum_{l=0}^{n-1}\left\vert l\right\rangle \left\langle
l\right\vert \ ,\ z\right]  \left(  1-\sum_{m=0}^{n-1}\left\vert
m\right\rangle \left\langle m\right\vert \right)  \left[  \bar{z}%
\ ,\ \sum_{l=0}^{n-1}\left\vert l\right\rangle \left\langle l\right\vert
\right] \label{S1}\\
&  =\theta^{-1}n\left\vert n-1\right\rangle \left\langle n-1\right\vert
.\nonumber
\end{align}
The topological charge is $Q=n$.

\section[ Discussions]{Discussions}

First we note that these solutions do not have the commutative counterparts.
In order to see this, we consider
\begin{equation}
W=\left(
\begin{array}
[c]{c}%
a^{-n}\bar{z}^{n}\prod_{l=1}^{n}\left(  \bar{z}z+l\theta\right)  ^{-1}\\
1
\end{array}
\right)  \label{w1}%
\end{equation}
for $Q<0$ and
\begin{equation}
W=\left(
\begin{array}
[c]{c}%
a^{-n}\left(  \prod_{l=1}^{n}\left(  \bar{z}z+l\theta\right)  ^{-1}\right)
z^{n}\\
1
\end{array}
\right)  \label{w2}%
\end{equation}
for $Q>0$, where $a$ is a real parameter. Taking the limit $a\rightarrow0$ in
the non-commutative case, $\Phi$ reduces to the non-singular configurations
(\ref{phi1}) and (\ref{phi2}) respectively. On the other hand, in the
commutative case, the same limit leads to
\begin{equation}
\Phi=\left(
\begin{array}
[c]{c}%
e^{-in\varphi}\\
0
\end{array}
\right)
\end{equation}
and
\begin{equation}
\Phi=\left(
\begin{array}
[c]{c}%
e^{in\varphi}\\
0
\end{array}
\right)  ,
\end{equation}
for $z\neq0$, where$\ z\equiv\left\vert z\right\vert e^{i\varphi}$. For $z=0$,
the limit leads to
\begin{equation}
\Phi=\left(
\begin{array}
[c]{c}%
0\\
1
\end{array}
\right)
\end{equation}
in both cases. There exists a discontinuous jump at the origin. Which shows
that in the commutative theory the configurations of $CP^{1}$ field $\Phi$
corresponding to $a\rightarrow0$ limit of (\ref{w1}), (\ref{w2}) are absent.
Thus, we have seen that the singular configurations in the commutative theory
become non-singular in the non-commutative case which can be attributed to
non-commutativity of the coordinates.

A relation to the $U\left(  2\right)  $ sigma model discussed in
Ref.\cite{Lechtenfeld:2001aw} is interesting. If we define $U=1-2P$ using the
projection operator $P$ of (\ref{Projection}), $U$ satisfies $U^{\dag}%
U=U^{2}=1$ and as such is a $U\left(  2\right)  $ field. Furthermore, the BPS
equation in $U\left(  2\right)  $ sigma model are expressed as (\ref{bps1-2})
and (\ref{bps2-2}) \cite{Lechtenfeld:2001aw}. From this viewpoint, our
solutions (\ref{P1}) and (\ref{P2}) could be considered as embedding of
abelian solutions into $U\left(  2\right)  $ sigma model. Actually, our
projector $P$ for $CP^{1}$ solitons is unitarily equivalent to%
\begin{equation}
P=\left(
\begin{array}
[c]{cc}%
1 & 0\\
0 & 0
\end{array}
\right)  -\left(
\begin{array}
[c]{cc}%
P_{n} & 0\\
0 & 0
\end{array}
\right)
\end{equation}
for $Q=-n$ and%
\begin{equation}
P=\left(
\begin{array}
[c]{cc}%
1 & 0\\
0 & 0
\end{array}
\right)  +\left(
\begin{array}
[c]{cc}%
0 & 0\\
0 & P_{n}%
\end{array}
\right)  ,
\end{equation}
for $Q=n$, where $P_{n}$ is the projector for abelian solitons of
ref.\cite{Lechtenfeld:2001aw}. Accordingly, the computations of energy for our
solitons could be reduced to those for the embedded abelian solitons. Which
makes it easy to verify $E=2\pi n$ by means of BPS equations.

Finally, a few words on the properties of the solutions (\ref{phi1}) and
(\ref{phi2}). It follows from (\ref{Projection}), that with the scalar
function $\Delta$ that commutes with $\bar{W}W$, the transformation
$W\rightarrow W\Delta$ does not change $\Phi$. In order to take the
$a\rightarrow0$ limit in (\ref{w1}) (\ref{w2}), we can use this invariance and
rewrite these as%
\begin{equation}
W=\left(
\begin{array}
[c]{c}%
\bar{z}^{n}\newline\\
a^{n}\prod_{l=1}^{n}\left(  \bar{z}z+l\theta\right)
\end{array}
\right)  ,\newline%
\end{equation}
and%
\begin{equation}
W=\left(
\begin{array}
[c]{c}%
z^{n}\\
a^{n}\bar{z}^{n}z^{n}+\sum_{m=0}^{n-1}\left\vert m\right\rangle \left\langle
m\right\vert
\end{array}
\right)  ,
\end{equation}
where
\begin{equation}
\Delta=a^{n}\prod_{l=1}^{n}\left(  \bar{z}z+l\theta\right)
\end{equation}
for (\ref{w1}) and
\begin{equation}
\Delta=a^{n}\bar{z}^{n}z^{n}+\sum_{m=0}^{n-1}\left\vert m\right\rangle
\left\langle m\right\vert
\end{equation}
for (\ref{w2}). Taking the limit of $a\rightarrow0$, the BPS soliton solutions
(\ref{w1}) (\ref{w2}) can be expressed in terms of $W$ as%
\begin{equation}
W=\left(
\begin{array}
[c]{c}%
\bar{z}^{n}\\
0
\end{array}
\right)  , \label{WV1}%
\end{equation}
and%
\begin{equation}
W=\left(
\begin{array}
[c]{c}%
z^{n}\newline\\
\sum_{m=0}^{n-1}\newline\left\vert m\right\rangle \left\langle m\right\vert
\end{array}
\right)  . \label{WV2}%
\end{equation}
The configurations (\ref{WV1}) and (\ref{WV2}) satisfy the BPS equations
(\ref{bps2-3}) and (\ref{bps1-3}) with
\begin{equation}
V=\bar{z}%
\end{equation}
and
\begin{equation}
V=z-\sqrt{n\theta}\left\vert n-1\right\rangle \left\langle n\right\vert ,
\end{equation}
respectively.

To summarize, we have considered solitons in the non-commutative $CP^{1}$
model and have found new solitons that do not have their counterparts in the
commutative theory. Further properties of these solutions will be examined
using more general approaches\cite{Hashimoto:2000kq}\cite{Hamanaka:2000aq}%
\cite{Hamanaka:2001dr} in a future publication.

\providecommand{\href}[2]{#2}\begingroup\raggedright

\end{document}